\documentclass[useAMS,usenatbib]{mn2e} 
\pdfoutput=1 
\usepackage{color}
\usepackage{amssymb}
\usepackage[sumlimits]{amsmath}
\usepackage{textcomp}
\usepackage{aas_macros}
\usepackage{epsfig}
\usepackage{multicol}
\usepackage{multirow}
\usepackage{graphicx}
\usepackage{booktabs,caption,fixltx2e}
\usepackage[flushleft]{threeparttable}
\usepackage{multirow}
\usepackage[sumlimits]{amsmath}
\usepackage[caption=false]{subfig}
\usepackage[normalem]{ulem}

\usepackage{soul}

\title[Modelling of the EBL]{Detailed modelling of the EBL along VHE $\gamma$-ray paths} 
\author[A. M. Kudoda and A. Faltenbacher]{A. M. Kudoda$^{1}$\thanks{E-mail:aymankudoda@gmail.com} and A. Faltenbacher$^{1}$\\
  $^{1}$School of Physics, University of the Witwatersrand, Johannesburg, 2050}
\begin{document}
\date{Accepted 2018 August 16. Received 2018 August 16; in original form 2018 May 31. }
\pagerange{\pageref{firstpage}--\pageref{lastpage}} \pubyear{2014}
\maketitle
\label{firstpage}
\begin{abstract}
  Interactions between the extragalactic background light (EBL) and very high energy $\gamma$ rays (VHE; $E > 10 \rm ~  GeV$) from cosmological sources alter their $\gamma$-ray spectrum. The stronger absorption of harder $\gamma$ rays causes a steepening of the observed $\gamma$-ray spectrum at the high energy tail which can be expressed by an increase of the power index. The effect provides a link between high energy astrophysics and the evolution of galaxies. In this work we develop a new hybrid EBL model by augmenting our previous analytic model with information from semi-analytic galaxy catalogues. The model allows us to study the $\gamma$-ray opacities of individual $\gamma$-ray paths through the (simulated) universe and evaluate the effect of local fluctuations in the EBL intensity along each path. We confirm an order of magnitude fluctuations in the local EBL (based on the cumulative light from model galaxy distributions within spheres of ~ $50 ~ h^{-1} \rm Mpc$). However, the effect of these fluctuations on the VHE $\gamma$-ray opacity is insignificant due to the overall very small contribution of the local EBL to the total EBL. We also investigate the effect which a galaxy crossing the line of sight of the $\gamma$-ray source may have on the VHE spectrum. We find that galaxies with stellar masses of $M >10^{11} \rm ~ M_\odot$ could have significant effect on the $\gamma$-ray absorption. It is unlikely that the observed variation in the spectral index is caused by the proximity of single galaxies to the $\gamma$-ray paths as these are extremely rare.
\end{abstract}
\begin{keywords}
  diffuse radiation -- dust, extinction -- gamma rays: observations -- stars: formation -- stars: fundamental parameters -- stars: luminosity function, mass function, semi-analytical model
\end{keywords}

\section{Introduction}
\label{sec:intro}
The extragalactic background light (EBL) is the integrated diffuse light from stars and active galactic nuclei (AGN) emitted through the history of the Universe redshifted to longer wavelengths due to cosmic expansion.  The spectrum of the EBL lies between the ultraviolet (UV) and far-infrared (FIR) with two distinct peaks: a first peak at $\sim 1 \rm ~ \mu m$ due to direct stellar emission; and a second peak at $\sim 100 \rm ~ \mu m$ caused by stellar light that has been absorbed and re-emitted  by the intra-galactic dust. Since the EBL carries information of the star formation history of the Universe, its measurements can be used to provide constraints on star formation models and the baryonic matter content of the Universe \citep{Dwek2001}.

Direct measurements of the EBL are difficult because of strong foreground contamination by galactic and zodiacal light.  Nevertheless, efforts have been made to provide upper and lower limits on the EBL using intensity measurements of the sky and  deep galaxy counts, respectively \cite[see][and references therein]{Dwek2001,Dwek2012}, and references therein). However, setting lower limits based on galaxy counts is less reliable at longer wavelength since unresolved galaxies become source of confusion. Furthermore, constraining the EBL in the $10-70\, \mu$m range is complicated by the emission from the interplanetary dust \citep{Kelsall1998}.

On the other hand, indirect measurements of the EBL can be obtained by studying the spectral attenuation of distant $\gamma$-ray sources. Cosmic $\gamma$-rays can be absorbed along the way through pair production, $\gamma+\gamma'\rightarrow e^+ + e^-$,  where $\gamma'$ indicates an EBL photon.  This process causes spectral steepening of blazars spectra in the very high energy (VHE; $E>10 \rm ~ GeV$) regime  \citep{Abramowski2013}. Since the attenuation depends on the $\gamma$-ray energy and the EBL density along the line of sight, it can be used to constrain the EBL. This method requires an understanding of the $\gamma$-ray source spectrum which is still a matter under discussion  \citep[e.g.,][]{Stecker1996, Dwek2005, Aharonian2006, Mazin2007, Albert2008}.  Thus, it is difficult to differentiate between the source-inherent effects and the signature of the EBL on the observed spectrum \citep{Mazin2007}.

\begin{figure}
  \begin{center}
    \includegraphics[width=0.5\textwidth]{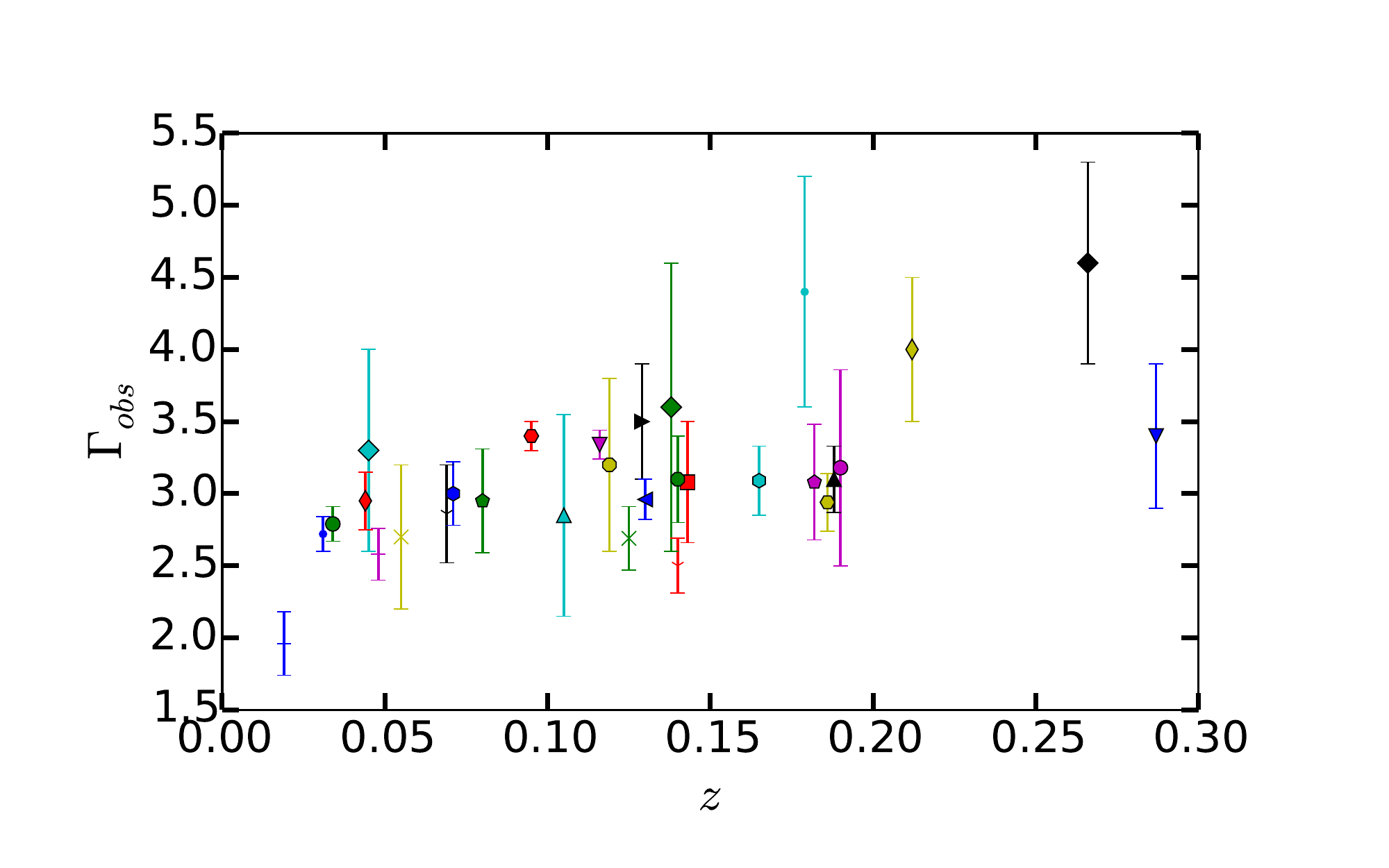}
    \caption{Observed spectral indices $\Gamma$ of HBL as a function of redshift (data obtained from table 1 in \protect\cite{Sinha2014}) excluding extreme HBLS and those with uncertain redshifts.}
    \label{fig:slo}
  \end{center}
\end{figure}

As illustration of the above we replicate a figure from \cite{Sinha2014} (Fig.~\ref{fig:slo}) which shows a positive correlation between the spectral index ($\Gamma$) of the high-frequency peaked BL Lac (HBL) sources and redshift, suggesting that more distant sources have a higher probability to interact with the EBL along the path leading to the steepening of the VHE spectra. The authors excluded extreme HBLs for homogeneity purposes which brings out more clearly the observed intrinsic systematic spectral hardening within this blazar class \citep[see also,][]{Ackermann2011}. However, even with this homogeneous sample there still is a large scatter in the $\Gamma$s for sources at the same redshift. Various models of blazars emission mechanisms indicate that blazars have a broad range of $\gamma$-ray peak frequencies, and therefore varying VHE spectral indices. In this work, we investigate a possible contribution from the EBL fluctuations to the scatter of the observed spectral indices.

Several models have been developed to derive an overall spectrum of the local EBL and its evolution. These models use different strategies to determine the evolution of the comoving luminosity density as a function of redshift. Most recent models can be classified into to four types: ``backward" evolution models, ``forward" evolution models, ``cosmic chemical" evolution models, and ``semi-analytic" models \citep{Dwek2001}.

Backward evolution models begin with the observed present day luminosity functions for galaxy populations in the local Universe and extrapolate them to higher redshift \cite[e.g.,][]{Malkan1998,Rowan-Robinson2001,Stecker2006}. Forward evolution models start with early structure formation scenarios to predict the galactic evolution forward in time \cite[e.g.,][]{Dwek1998,Razzaque2009,Finke2010,Kudoda2017a}.  Cosmic chemical evolution models consider basic galaxy ingredients such as gas, metallicity and radiation content and follow their  evolution in a large comoving volume element \cite[e.g.,][]{Pei1995,Pei1999}. Semi-analytic models apply physically motivated descriptions and recipes to simulate galaxy formation and evolution  \cite[e.g.,][]{Kauffmann1993, Cole1994, Somerville1999,Gilmore2012}. Semi-analytic models are a more challenging approach, but they provide a better insight into the complex physical processes involved in the production of the EBL.

The standard $\gamma$-ray absorption models assume spatially homogeneous EBL densities. However, the spatial galaxy distribution is inhomogeneous on small scales, thus one expects the EBL to fluctuate to a certain degree as well. To what extent fluctuations affect the $\gamma$-ray transmissivity is not fully understood. The effect of inhomogeneous EBL densities on $\gamma$-ray opacities has been explored by \cite{Furniss2015,Kudoda2015,Kudoda2017a} and \cite{Abdalla2017}. \cite{Kudoda2017a} (hereafter referred to as KF17) predicted maximum changes of $\pm 10 \%$ in the $\gamma$-ray transmissivity, which is in agreement with the results reported by \cite{Furniss2015} and \cite{Abdalla2017}. A change of $\pm 10 \%$ in the $\gamma$-ray transmissivity, however, translates into only marginal differences in the power law slopes of the absorbed $\gamma$-ray spectra ($\lesssim \pm 1\%$).

In KF17,  we developed a purely analytical framework for computing the EBL fluctuation based on the phenomenologically determined variance of the SFR. In this work we develop a new technique to model the EBL (which may be adaptable for other problems as well) by combining the forward and semi-analytic approaches in order to study the impact of the local environment on the overall EBL spectrum and its effect on the $\gamma$-ray opacity of the Universe.  The forward approach, borrowed from KF17, will be used to model the isotropic contribution to the local EBL from galaxies at large distances. While, the  semi-analytical galaxy catalogues from the Millennium database \citep{Lemson2006,Guo2013} will be used to create a light cylinder through the Universe to simulate the $\gamma$-ray path with a more realistic distribution of galaxies close to the line-of-sight.  Combining the two approaches allows us to model $\gamma$-rays on individual paths through a locally inhomogeneous EBL.

The structure of the paper is set as follows.  We detail the new EBL model in Section \ref{sec:mod}. The results of the EBL model and comparison with our previous EBL model are presented in Section \ref{sec:res}. The calculations of the $\gamma$-ray opacity are discussed in Section \ref{ssc:opa}.  The case of a galaxy encounter with the $\gamma$-ray path is shown Section \ref{ssc:gal}. Finally, we summarize and give a brief conclusion in Section \ref{sec:con}.  In this work we used the WMAP7 cosmological parameters  ($\Omega_m$, $\Omega_b$, $\Omega_\Lambda$, $h$) = ($0.272$, $0.0455$, $0.728$, $0.701 \rm ~ km/s/Mpc$).

\section{Modelling the EBL}
\label{sec:mod}
This work focuses on the effect of the stochasticity of local EBL fluctuations on the absorption of VHE $\gamma$-rays. We employ a semi-analytical galaxy catalogue to model the local EBL fluctuations. The spatial information of semi-analytical galaxy positions allow us to investigate the $\gamma$-ray absorption along individual $\gamma$-ray paths and evaluate the effects of different environments (EBL energy density along the line-of-sight).  Since the Universe can be considered homogeneous on scales $\gtrsim 100 \rm ~ Mpc$, it implies that the EBL originating for sources with large distances to the $\gamma$-ray paths can be considered as homogeneous. However, if the $\gamma$-rays pass over- or under-dense structures such as galaxy clusters or voids, the EBL is expected to take on higher or lower than the average values. Consequently, we divide the computation of our EBL model into two parts: the local EBL contribution (within  a radius $<50\,h^{-1} \rm ~ Mpc$)  derived directly from the semi-analytical galaxy positions and brightness; and the contribution from a homogeneous, analytically modelled, EBL background ($> 50\,h^{-1} \rm ~ Mpc$).

\subsection{The local EBL contribution}
The EBL differential specific flux at wavelength $\lambda_0$, dF$_\nu(\lambda_o)$,  arriving from sources contained by a comoving volume element dV$_c$(z) at redshift $z$ and wavelength $\lambda$ can be computed as \cite[e.g.][]{Mo2010}
\begin{align}\label{eq:Fnu}
  \text{dF}_\nu(\lambda_o) = (1+z) \frac{\mathcal{L}_\nu(\lambda,z) \text{d}V_c(z)}{4\pi d_L(z)^2},
\end{align}
where $\mathcal{L}_\nu(\lambda, z)$ and $d_L$  are the comoving specific luminosity density of the sources and their luminosity distance, respectively. Since this formula is used only for distances $\leq 50\,h^{-1} \rm ~ Mpc$ here we ignore the factor $(1+z)$ for the computation of the local EBL contribution.   

\begin{figure*}
  \begin{center}
    \includegraphics[width=1\textwidth]{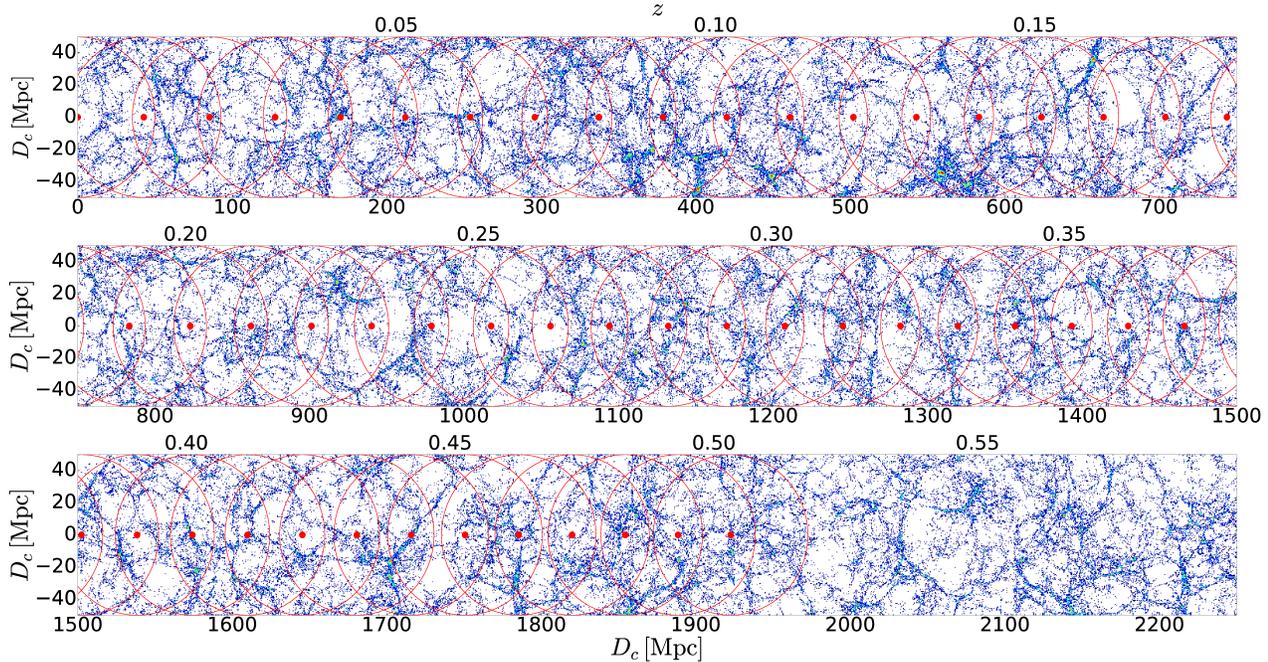}
    \caption{Example of the galaxy distribution within a cylinder enclosing an individual $\gamma$-ray path with comoving radius of $50\,h^{-1} \rm ~ Mpc$ from $z=0$ up to $z=0.5$. The red points in the middle of the cylinder represent the locations where the local EBL density from a sphere of radius $50 ~ h^{-1} \rm Mpc$ (red circles) was computed. Note that here we are plotting a $10 \rm ~ Mpc$ slice of the cylinder.}
    \label{fig:cyl}
  \end{center}
\end{figure*}

\begin{figure}
  \begin{center}
    \includegraphics[width=0.5\textwidth]{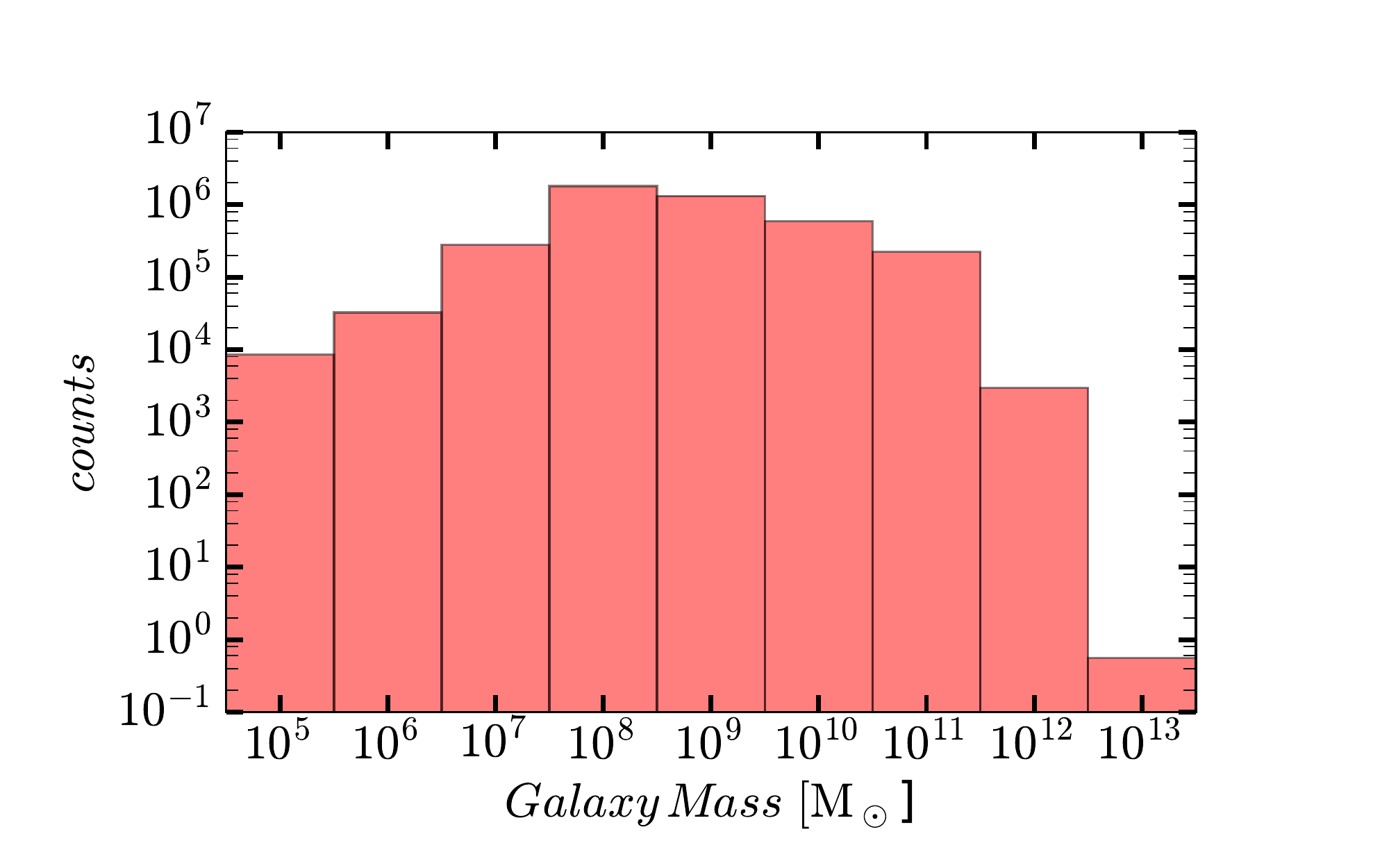}
    \caption{The mean galaxy stellar mass distribution within all the $961$ cylinders considered. The volume of a cylinder is ~$15.7\times 10^6$ ($h^{-1}\rm ~ Mpc)^3$.}   
    \label{fig:msd}
  \end{center}
\end{figure}

\subsubsection{The light-cone cylinder}
\label{ssc:lc}
To investigate the impact of the ambient luminous structures on different $\gamma$-ray paths we use the semi-analytical galaxy catalogue MR$7$ \citep{Lemson2006,Guo2013} and construct a rectangular volume with two dimensions coincident with two edges of the simulation cube and the third dimension extending into redshift space. In essence this is a rectangular section of the cosmic light cone. The MR$7$ data provides $62$ snapshots of a $500^3\,h^{-1}$ Mpc$^3$ comoving simulation volume from redshift $z=50$ up to $z=0$.  We draw a total of $961$ parallel $\gamma$-ray paths (with cylinders with a radius $R=50\,h^{-1} \rm ~ Mpc$ around them) separated by $\sim 15 \rm ~ Mpc$ through the light-cone starting from $z=0$ up to $z=0.5$ ($\simeq 2000\,h^{-1} \rm ~ Mpc$).

In order to avoid periodic repetition of structures within the light cones we randomly rotate the simulation cubes before stitching them together.  More precisely we cut a slice from each snapshot with a thickness equal to the comoving distance between two consecutive snapshots and place the slices at the comoving distance corresponding to their redshifts.  Figure \ref{fig:cyl} shows an example of the resulting cylinders, where the red dots represent the chosen points (separated by $dz=0.01$) to calculate the light intensity from the local galaxies located in a sphere with radius of $50 ~ h^{-1} \rm Mpc$ (plotted as red circles in the figure). The mean distribution of the stellar mass component of the galaxies in all cylinders is shown in Fig. \ref{fig:msd}. The stellar mass of most galaxies in the cylinders lie between $10^7$ to $10^{11} \rm ~ M_\odot$.

\begin{figure}
  \begin{center}
    \includegraphics[width=0.5\textwidth]{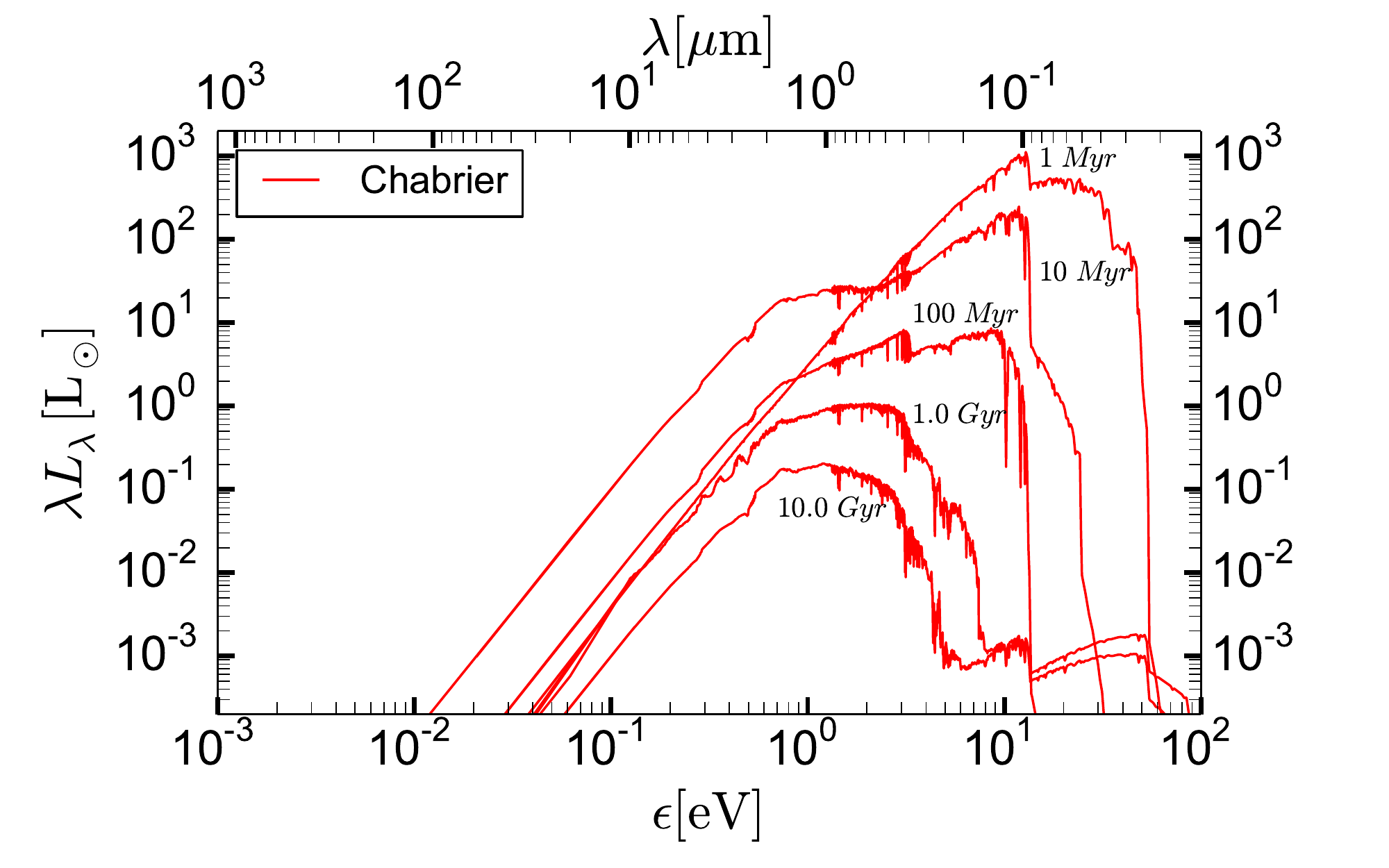}
    \caption{The \protect\cite{Bruzual2003} dust free spectrum using Padova 1994 evolutionary tracks with Chabrier IMFs at different ages.}
    \label{fig:bc03}
  \end{center}
\end{figure}

To assign a spectrum to every galaxy in the cylinders we use the galaxy properties: age, metallicity, and stellar mass (all listed in the MR7 catalogue) together with the stellar population synthesis library BC03 \citep{Bruzual2003}, assuming each galaxy has only a single stellar population. BC03 use Padova 1994 stellar evolutionary tracks and \cite{Chabrier2003} initial mass function to determine high resolution spectral evolution of stellar populations with different metallicities and ages. Figure \ref{fig:bc03} shows the dust free spectrum  normalized with solar luminosity ($\rm L_\odot$) at different ages.  

\begin{figure}
  \begin{center}
    \includegraphics[width=0.5\textwidth]{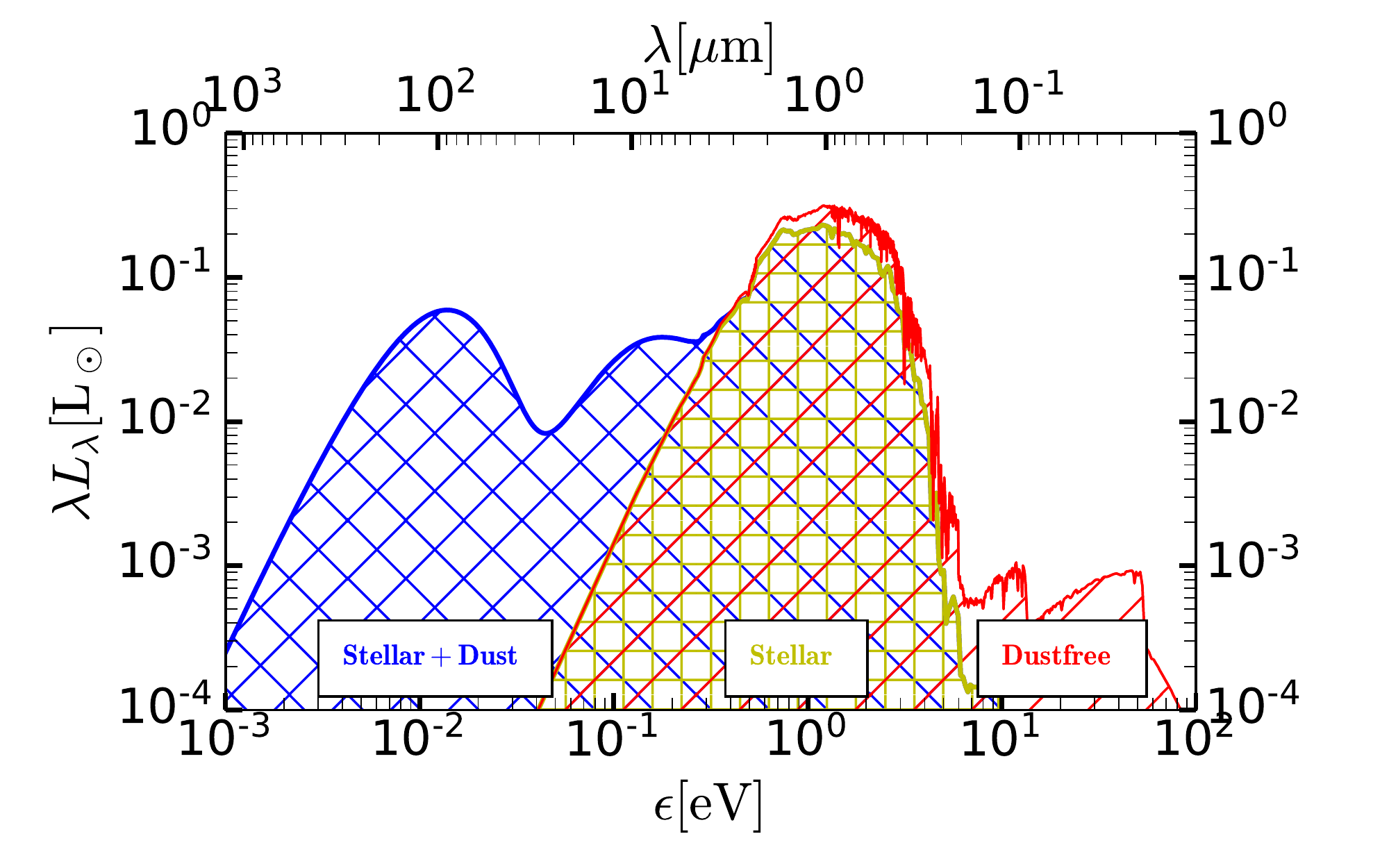}
    \caption{Illustration of the composition of a galaxy spectrum normalized with solar luminosity. The dust free model at $5$ Gyr from \protect\citep{Bruzual2003} in red. The yellow represents the stellar emission, while the blue lines is the total spectrum (stellar + dust extinction).}
    \label{fig:sd}
  \end{center}
\end{figure}

To get the total spectrum of the individual galaxies and estimate the absorbed light by the dust  we apply an empirical formula for the light escape fraction from \citep{Razzaque2009} on the BC03 spectra and generate the stellar component spectrum. The dust content is modelled by three components large dust grains; small dust gains; and polycyclic aromatic hydrocarbons (PAHs) with the temperatures $40$, $70$ and $450  \rm ~ K$ respectively \citep{Finke2010}.  In order to get the dust spectrum we integrate the total energy density of the photons absorbed by the dust and distribute it over three blackbodies representing the three dust components with the fractions $0.60$, $0.05$, and $0.35$, respectively. The full spectrum is then obtained by combining the stellar and the dust spectrum. Figure \ref{fig:sd} shows dust free, stellar and dust+stellar SEDs of the stellar population normalized to $\rm M_\odot$ at an age of $5$ Gyr and solar metallicity. 

In order to calculate the local contribution of the EBL at the sampling-points (the red dots in Fig. \ref{fig:cyl}) we derive the EBL intensity by utilising Eqn. \ref{eq:Fnu} and add the contribution of the light from all galaxies within the $50 ~ h^{-1} \rm Mpc$ comoving sphere. In this way we produce the local EBL spectrum for all sampling-points along the line-of-sight.

\begin{figure}
  \begin{center}
    \includegraphics[width=0.5\textwidth]{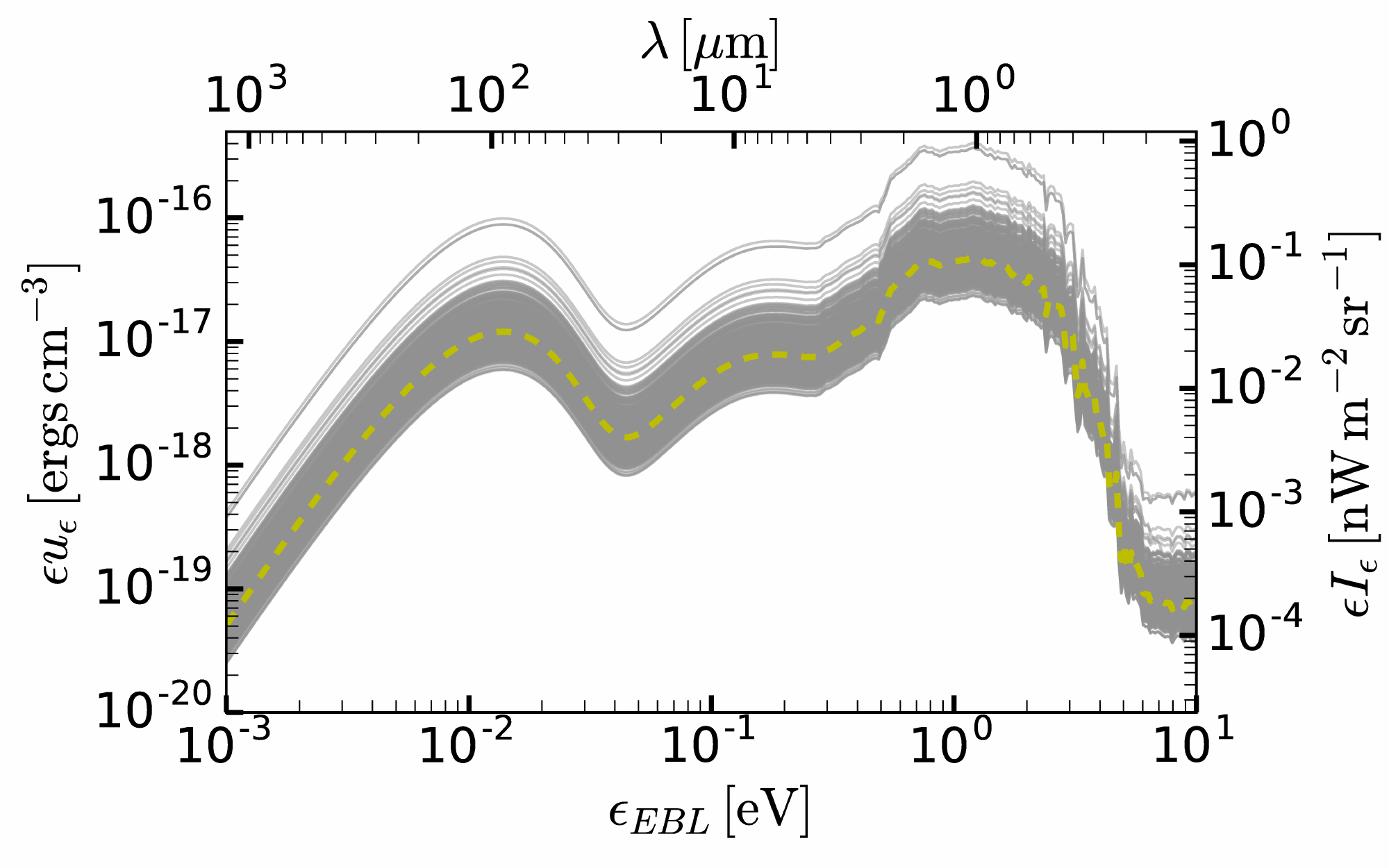}
    \caption{The local EBL spectra in different spheres at $z=0$ (grey lines). The yellow dashed line represents the median local EBL in one sphere as an example.  The observed fluctuations between different spheres can be explained by the variation in the number of galaxies or large structures in each sphere and their distances from the centre of the sphere. In other words, the lines reflect the variation of the distance of the large structures from the line-of-sight.} 
    \label{fig:sph}
  \end{center}
\end{figure}

\begin{figure}
  \begin{center}
    \includegraphics[width=0.5\textwidth]{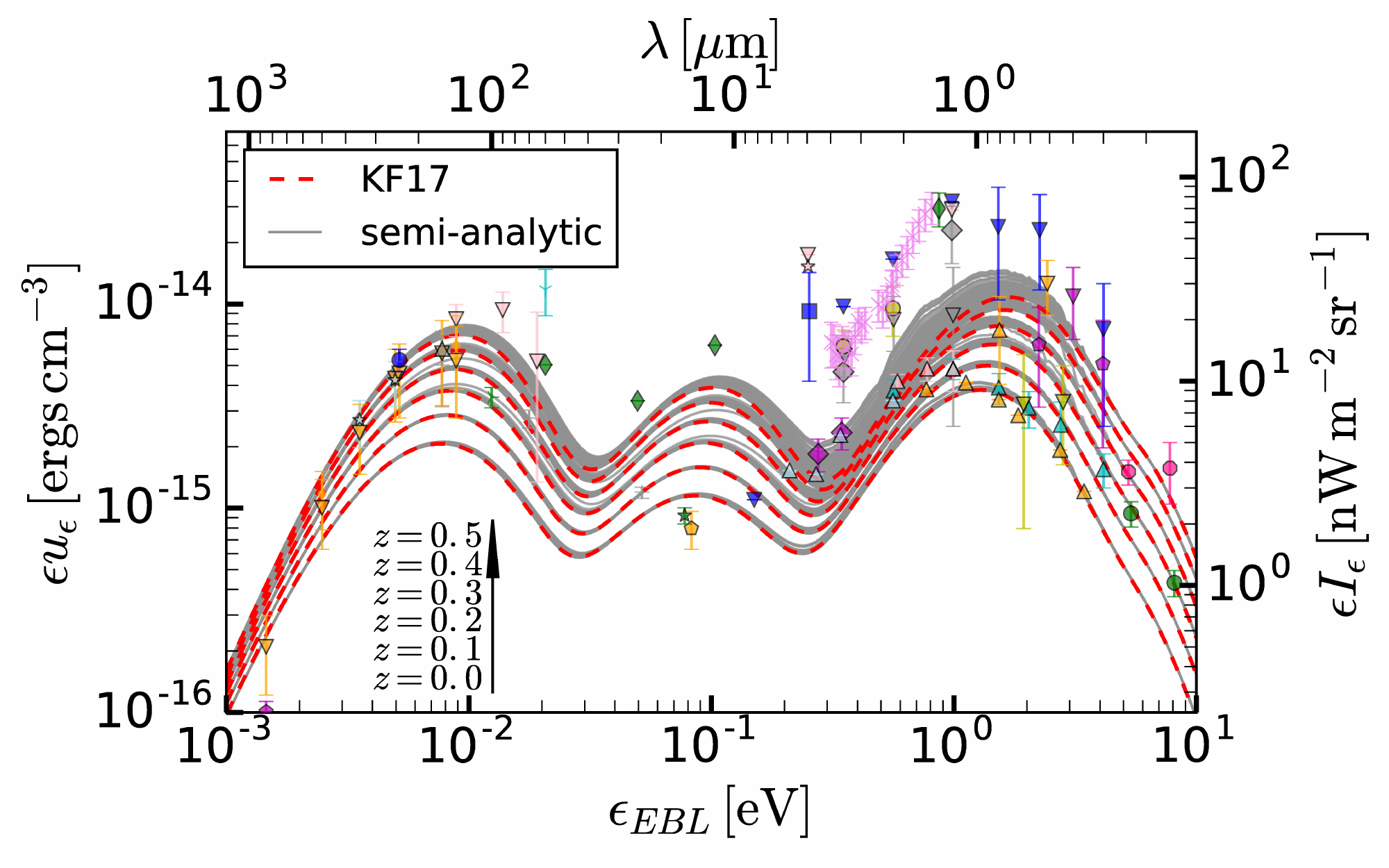}
    \caption{The total EBL intensity at different redshifts (grey lines). The red dash-lines represent the analytic KF17 model.  The symbols represent EBL measurements: green circle \protect\citep{Xu2005}; red octagon \protect\citep{Gardner2000}; blue triangle down \protect\citep{Bernstein2007}; cyan triangle up \protect\citep{Totani2001}; magenta triangle down \protect\citep{Mattila1990}; yellow triangle down \protect\citep{Matsuoka2011}; orange triangle down \protect\citep{Dube1979}; Grey triangle down \protect\citep{Levenson2007}; pink triangle up \protect\citep{Keenan2010}; magenta diamond \protect\citep{Ashby2013}; blue square \protect\citep{Arendt2003}; orange pentagon \protect\citep{Hopwood2010}; green star \protect\citep{Teplitz2011}; Grey plus \protect\citep{Bethermin2010}; cyan triangle down \protect\citep{Finkbeiner2000}; pink triangle down \protect\citep{Matsuura2011}; green triangle up \protect\citep{Berta2011}; Orange triangle down \protect\citep{Fixsen1998}; Grey triangle down \protect\citep{Penin2012}; red star \protect\citep{Bethermin2012}; Orange dot \protect\citep{Odegard2007}; blue circle \protect\citep{Zemcov2010}; the yellow x \protect\citep{Matsumoto2005}; green diamonds \protect \citep{Kashlinsky2000}. }
    \label{fig:EBL}
  \end{center}
\end{figure}

\begin{figure}
  \begin{center}
    \includegraphics[width=0.5\textwidth]{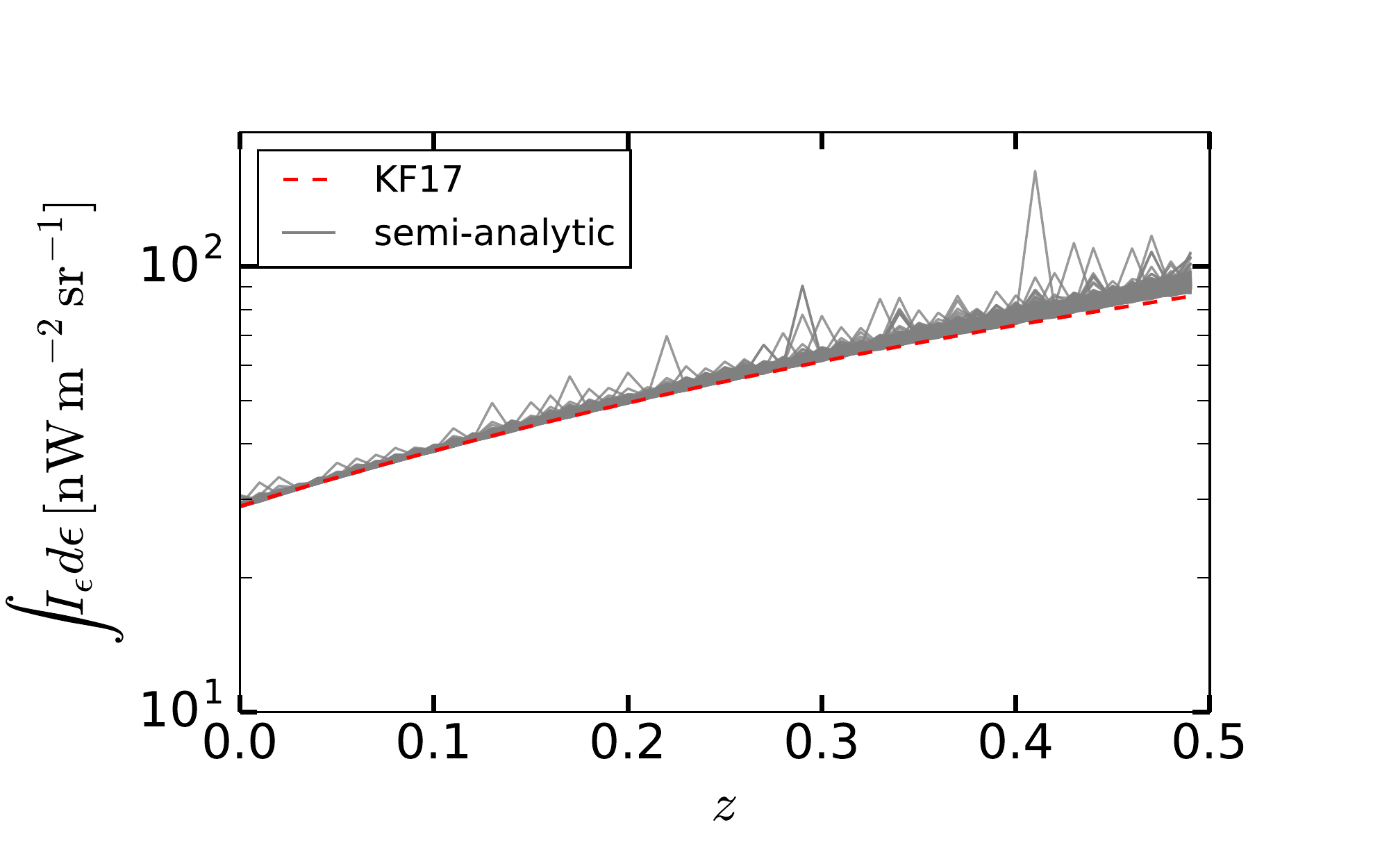}
    \caption{The grey lines show the total EBL intensity along the various $\gamma$-ray paths (961 cylinders) starting from $z=0$ up to $z=0.5$. The red dash-doted line is the pure analytic approach KF17 from \protect\cite{Kudoda2017a}, the red dashed line shows an example of the integrated total EBL intensity in one cylinder}.
    \label{fig:EBLI}
  \end{center}
\end{figure}

\subsection{The homogeneous EBL background}

For the homogeneous EBL background (beyond 50 Mpc) we use our previous EBL forward model KF17 \citep{Kudoda2017a}, considering only the contribution form main-sequence stars to the EBL stellar component. The EBL at $z = z_1$ is derived by integrating the contributions from stars of all masses formed throughout the history of the universe from $z=10$ to $z = z_1$. The number of photons $N$ emitted from a star with mass $M_\star$ at specific energy interval $d\epsilon$ and time $dt$ is modelled as black-body radiation at a given temperature $T_\star$ and radius $R_\star$ from the spherical surface of the star.  An initial mass function (IMF, $\xi(M)$) is then employed to model the mass distribution of the stars. A further element for the integration of the EBL is the global SFR ($\psi(z)$),
\begin{align}\label{eq:R}
  \frac{dN(\epsilon,z=z_i)_{star}}{d\epsilon dV}= & \,\mathcal{N} \int^\infty_{z=z_i} dz'' \left|\frac{dt}{dz''}\right| \psi(z'') \int^{M_{\text{max}}}_{M_{\text{min}}} dM \xi(M) \nonumber\\ 
  &\,\times\int^{z''}_{\text{max}\{z=0,z_d(M,z')\}}dz' \left|\frac{dt}{dz'}\right|\nonumber\\ 
  &\,\,\,\, f_{\rm esc}(\epsilon') \frac{dN(\epsilon',M)}{d\epsilon'dt}(1+z'), 
\end{align}
where $\mathcal{N}$ is the normalisation factor for the IMF $\xi(M)$, $M_{min}$ and $M_{max}$ are $0.1 \rm ~ M_\odot$  and $100 \rm ~ M_\odot$, $\epsilon'=\epsilon(1+z')$ is the redshifted energy of the EBL photons, $z_d$ is the redshift when the star evolves away from the main-sequence, and $\frac{dN(\epsilon',M)}{d\epsilon'dt}$ is the total number of emitted photons per time per energy intervals. The averaged photon escape fraction from a galaxy, $f_{\rm esc}(\epsilon')$, is given by an empirical fitting function adapted from \cite{Driver2008}. The fraction of photons generated in stars which does not escape is $1-f_{\rm esc}(\epsilon')$.

To compute the dust component $(\frac{dN(\epsilon,z=0)_{dust}}{d\epsilon dV})$ we followed the recipe discussed in the previous section by replacing the third integral in Eq. \ref{eq:R} with:

\begin{align}\label{eq:Rplus}
  &\times\int^{z''}_{\text{max}\{z=z_1,z_d(M,z')\}}dz' \left|\frac{dt}{dz'}\right| (1+z') \sum_{n=1}^3 {{\Omega_n(M) \epsilon'^2}\over{\exp(\epsilon'/kT_n)}} 
\end{align} 

where $n$ is an arbitrary index, $\Omega_n$ are the scaling factors of the fraction of star-light re-radiated by a particular black-body at temperature $T_n$. Combining the stellar and dust components we can obtain the full spectrum. 

\subsection{The total EBL spectrum}

Finally, we derive the total EBL intensity by adding the local contribution at $z_i$ to the homogeneous contribution $z_i-0.01$, which is approximately the redshift corresponding to a $50 ~ h^{-1} \rm Mpc$  comoving distance (the radius of the sphere used to compute the local EBL contribution). The comoving radius of the observable Universe is roughly $14.26 \rm ~ Gpc$  which is roughly two orders of magnitude larger than the radius of the sphere used for the computation of the local EBL contribution. Therefore, one would expect the light intensity of the homogeneous background to be roughly two orders of magnitude larger than the intensity form the local contribution. Since the analytic background model only considers main-sequence stars and assumes a redshift independent IMF and dust absorption to determine the EBL density, this results in an EBL density lower than the observed one.

The inclusion of the post-main-sequence stars in the local EBL increases the EBL density slightly compared to the analytical model but this contribution is not sufficient to outweigh the shortfall of EBL contribution from the analytical model.

\begin{figure}
  \begin{center}
    \includegraphics[width=0.5\textwidth]{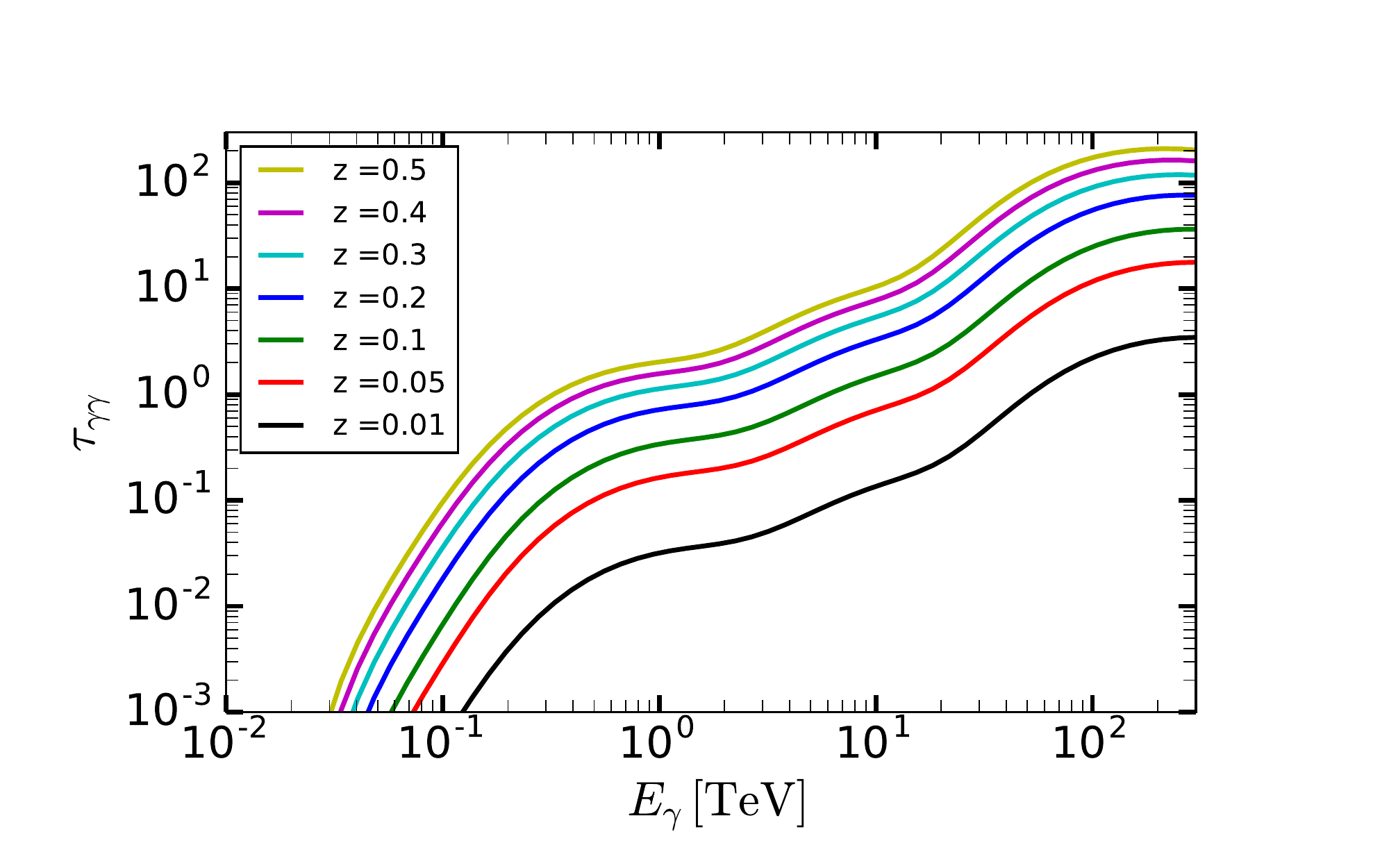}
    \caption{The $\gamma$-ray opacity averaged over all $\gamma$-ray paths at different redshifts.}
    \label{fig:opc}
  \end{center}
\end{figure}

\begin{figure}
  \begin{center}
    \includegraphics[width=0.5\textwidth]{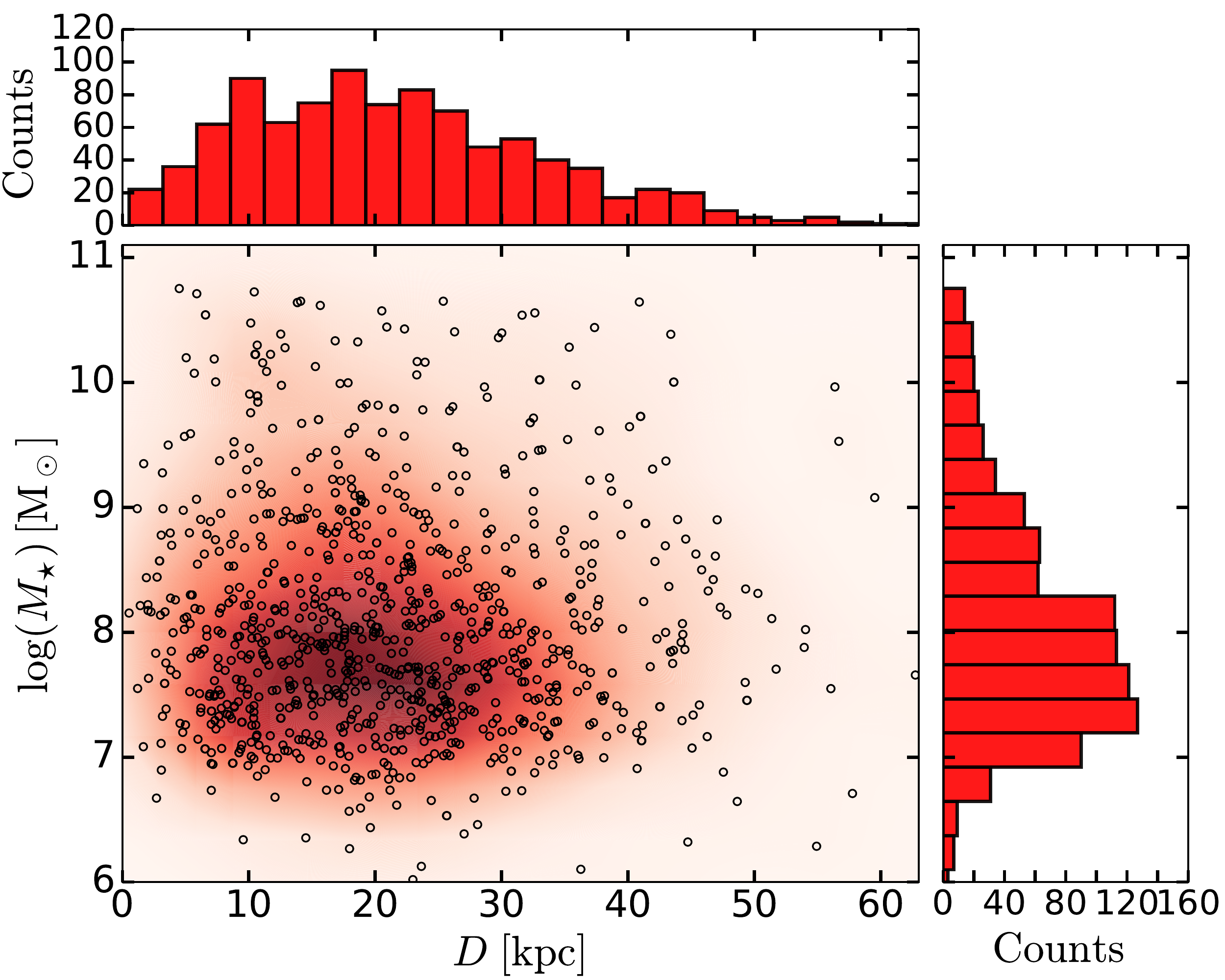}
    \caption{The distribution of the distances and stellar masses of the galaxies closest to the individual $\gamma$-ray path (one galaxy per $\gamma$-ray path).}
    \label{fig:dist}
  \end{center}
\end{figure}

\begin{figure*}
  \begin{center}
    \includegraphics[width=0.95\textwidth]{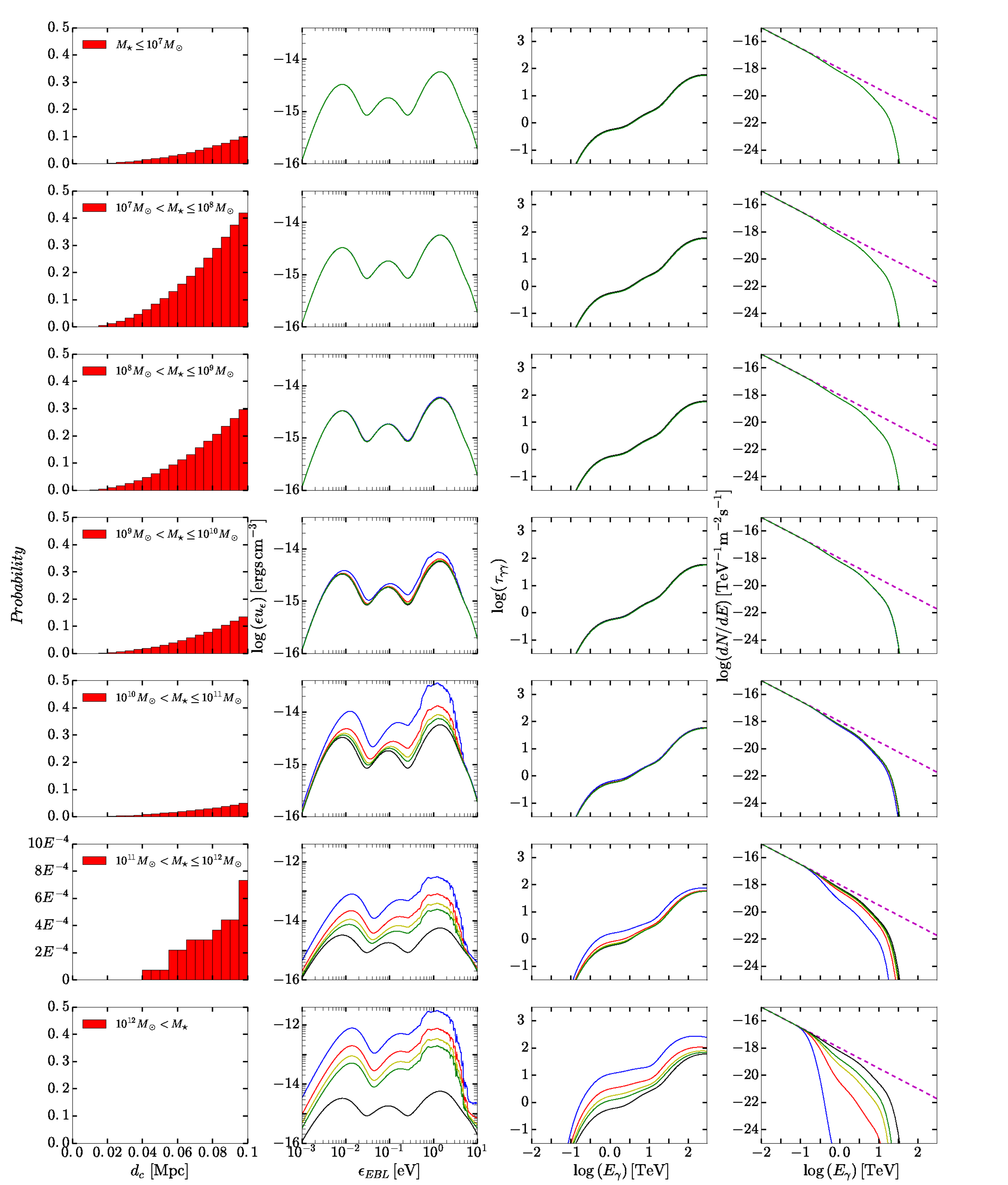}
    \caption{Effect of close encounter between the $\gamma$-ray path and an intervening galaxy on the EBL energy density, the $\gamma$-ray opacity and the VHE $\gamma$-ray spectrum. The intervening galaxy is assumed to be located at a distance corresponding to $z=0.15$. The first column from the left shows the probability distribution of finding a galaxy at a specific radius (derived from the relative positions of the semi-analytical galaxy population and the assumed $\gamma$-ray paths), where the rows from top to bottom represent different stellar mass bins. The second column shows the impact on the EBL energy density when we insert a galaxy at $20$, $40$, $60$ and $80 \rm ~ kpc$ from the line-of-sight represented by blue, red, yellow and green lines, respectively. For comparison the EBL energy density at $z=0.15$ without intervening galaxy is shown by the black line. Assuming the $\gamma$-ray source at $z=0.16$, the third column gives the opacities corresponding to the different EBL spectra shown in the second column. The last column presents the impact of the altered opacities on the shape of VHE spectra with intrinsic power law index of $\Gamma = -1.5$ at $z=0.16$. The dashed magenta line indicates the intrinsic VHE spectrum without EBL absorption.}
    \label{fig:mas}
  \end{center}
\end{figure*}

\section{Results}
\label{sec:res}
In this section we first present the results of our hybrid EBL model and a comparison with our previous pure forward analytic EBL model. We then discuss the fluctuations of the $\gamma$-ray opacity using the EBL density along the different $\gamma$-ray paths.

\subsection{EBL spectra}
We apply the method described in Sec. \ref{sec:mod} to calculate the EBL energy density, i.e., model the contribution of the local EBL and the homogeneous background separately. Figure \ref{fig:sph} shows the local EBL spectra from the $961$ spheres at $z=0$ as grey lines. The median of the spectra is indicated by the yellow dashed-line. We observe an EBL intensity variation of one order of magnitude between the spheres which is a result of the different galaxy content within the individual $50\,h^{-1}\rm ~ Mpc$ spheres. We observe a steep decline of the local EBL intensity above $4 \rm ~ eV$. This is due to the prevalence of galaxies with old stellar populations (see Fig. \ref{fig:bc03}).

In Fig. \ref{fig:EBL} we show the hybrid total EBL spectra (the summation of the local contribution and the homogeneous background) within all light-cone cylinders at different redshifts as grey lines and compare it with the EBL intensity from our previous pure analytic model KF17, displayed as red dashed lines and observations given by the symbols. The total EBL intensity increases with redshift which is a consequence of the increasing number density of galaxies in physical coordinates. The total EBL spectrum at $z=0$ is $\pm 1 \%$ the spectrum in the KF17 model. This is the result of the more comprehensive simple stellar population (SSP) evolution model used to determine the emission from semi-analytical galaxies employed for the estimation of the local EBL here.

The magnitudes of the total EBL intensity derived from our model are at the lower limit of the observations but roughly in agreement with the data points by \cite{Totani2001} and \cite{Fixsen1998}. The low model EBL intensities are caused by the over-simplified treatment of the stellar component employed for the calculation of homogeneous background. Since our main focus is the impact of the EBL fluctuations on $\gamma$-ray opacity an exact matching of the observed EBL magnitudes is not necessary.

The fluctuations in the total EBL, \i.e. the variation of the magnitudes of the model EBL spectra for a specific redshift (bundles of grey lines in Fig.~\ref{fig:EBL}), are small compared to the fluctuations in the local EBL (Fig.~\ref{fig:sph}) because the isotropic background is more than $\sim 100$ times brighter than the local contributions. The fluctuations between $\sim 0.004-4 \rm ~ eV$ are more visible as a result of the local EBL spectrum shape (see Fig. \ref{fig:sph}) where older galaxies dominate the spectrum.

Figure \ref{fig:EBLI} shows the evolution of the total integrated EBL spectrum ($\int I_\epsilon d \epsilon$ from $0.001$ to $10 \rm ~ eV$) as a function of redshift. Each grey line displays the integrated EBL intensity along an individual $\gamma$-ray path. As expected the amplitude of the fluctuations here also increases with redshift for the same reason mentioned above. The red dash-line depicts the KF17 model and indicates good agreement between the new semi-analytical approach and the KF17 model.

\subsection{Gamma-ray opacity}
\label{ssc:opa}
Based on the total cross section $\sigma$ for pair production in the photon-photon interaction described in \citep[cf.][]{Gould1967}, we calculate the $\gamma$-ray opacity $\tau_{\gamma\gamma}$ as a function of the $\gamma$-ray energy $E_\gamma$ and redshift, $z$ \citep[cf.][]{Razzaque2009}:   

\begin{align}\label{eq:opt}
   \tau_{\gamma\gamma}(E_\gamma,z) = c \pi r_0^2\left(\frac{m_e^2c^4}{E_\gamma}\right)^2 &\int^z_0 \frac{dz_1}{(1+z_1)^2}\left| \frac{dt}{dz_1}\right| \nonumber\\ 
    \times &\int^\infty_{m_e^2c^4/E_\gamma(1+z_1)} d\epsilon_1\frac{u_{\epsilon_1}}{\epsilon_1^3} \overline{\varphi}[s_0(\epsilon_1)]\ ,
\end{align}
where $r_0$ is the classical electron radius, $s = E_\gamma (1+z_1) \epsilon_1 / 2m_e^2c^4$ is the center-of-momentum energy square, $\varphi[s_0(\epsilon_1)] = \int^{s_0(\epsilon)}_1 \frac{2s\sigma(s)}{\pi r_0^2}ds$ and  $s_0 = E_\gamma(1+z_1)\epsilon_1 / m_e^2c^4$.

Using the above integral we determine the optical depth as function of $\gamma$-ray energy for different redshifts. Figure \ref{fig:opc} shows the $\gamma$-ray opacity in the range $0.01-300 \rm ~ TeV$  for the $961$ $\gamma$-ray paths as a function of the $\gamma$-ray source redshift. The opacity increases rapidly above $1\rm ~ TeV$ , implying that $\gamma$-ray photons at the tail of the spectrum are most attenuated.  The overall opacity also increases with redshifts showing that $\gamma$ rays from distance sources experience more attenuation along their path.  In addition, the resulting opacities for all $\gamma$-ray paths overlap each other at all redshifts.  This means that the small  EBL fluctuations along the $\gamma$-ray path, shown in Fig. \ref{fig:EBLI} as grey zigzag lines, are smoothed out over redshift suggesting an average opacity along vast distances. We can also say that the different structures  along the various $\gamma$-ray paths have no impact on the total EBL density and the consequent opacity over long distances.  The absence of fluctuations on the total EBL densities and opacities along all possible $\gamma$-ray paths simulated here agrees with the purely analytical results in KF17 where the variation in the EBL and the opacity was only $\pm 1 \%$.

In conclusion, we find that the $\gamma$-ray opacities along the different $\gamma$-ray paths simulated here do not show significant fluctuations. 

Because of the finite number of $\gamma$-ray paths used here in combination with the relatively low number densities of massive galaxies (see Fig. \ref{fig:dist}), which substantially enhance the local EBL photon density, we were not able to explore close encounters between massive galaxies and the $\gamma$-ray paths in a comprehensive manner.  Therefore, in the following section we investigate close encounters of galaxies with $\gamma$-ray paths by manually moving a model galaxy into the line-of-sight.

\section{The impact of close galaxy encounters}
\label{ssc:gal}
In this section we study the effect of galaxy encounters with the $\gamma$-ray path on EBL energy density, opacity and $\gamma$-ray spectral index.  To obtain an insight into the frequency of close galaxy - $\gamma$-ray encounters we select for each $\gamma$-ray path the galaxy with the minimal perpendicular distance to the line-of-sight. Figure \ref{fig:dist} shows the mass distribution of the $961$ selected galaxies as a function of distance from their corresponding $\gamma$-ray paths. The stellar mass of most of these galaxies fall in the range of $10^7-10^8 \rm ~ M_\odot$ between $10$ to $30\rm ~ kpc$ from the line-of-sight. No galaxy with mass $\gtrsim 10^{11} $ is found in the distance less than $60 \rm ~ kpc$ form the $\gamma$-ray path. This is consistent with the overall stellar mass distribution shown in Fig. \ref{fig:msd} which indicates a stellar mass range of the model galaxies from $10^7$ to $10^{10} \rm ~ M_\odot$. In general, massive galaxies are rarely found very close to the $\gamma$-ray paths simulated here.

However, one also has to consider the gravitational lensing effect caused by an intervening galaxy which counteracts the steepening of the VHE $\gamma$-ray spectrum if the line of sight goes directly through the visible part of the intervening galaxy \citep{Barnacka2014}. The deciding factor here is the impact relative to the Einstein radius $r_E$, which characterises the strength of the gravitational lensing. The Einstein radius depends on the mass of the  encountering galaxy, the angular distances from the $\gamma$-ray source to the observer, from the $\gamma$-ray source to the encountering galaxy, and from the encountering galaxy to the observer. The lensing effect is significant if the $\gamma$-ray passes through one $r_E \sim 5$ kpc of the intervening galaxy. Therefore, in this work we choose the distance between the $\gamma$-ray path and the intervening galaxy to be $\gtrsim 20$ kpc so that we cover the effect of the encounter outside the lensing radius.

In order to quantify the effect of a close encounter we now insert a galaxy close to the $\gamma$-ray path manually with a very simple model. In principle, this will allow us to vary the distance, mass and redshift of the encountering galaxy, and investigate their importance.   Figure \ref{fig:mas} summarises the effects of the presence of a galaxy near to the line-of-sighs at $z=0.15$ with the $\gamma$-ray source is at $z=0.16$. There are no changes on the opacities if we change the encountering galaxy's redshift. Each row in the figure represents  different stellar mass bins starting from  $M<10^7 \rm ~ M_\odot$ at the top to $M>10^{13} \rm ~ M_\odot$ at the bottom with a logarithmic mass bin. The first column shows the probability of finding a galaxy within a given perpendicular distance from the line-of-sight.  To compute the probability we summed all galaxies at certain mass and radial bins along all  line-of-sights and divide them by the total number of galaxies inside a maximum radius of $0.1 \rm ~ Mpc$.  Using the total number of galaxies for normalization instead of the number of galaxies in each mass bins separately enables us to easily compare the relative radial probability  distribution of the galaxies in each mass bins.  The second column displays the impact of a close galaxy encounter with the line-of-sight on the EBL at different distances ($20$, $40$, $60$, $80 \rm ~ kpc$). The third column gives the effect on the corresponding $\gamma$-ray opacity due to the EBL, and the last column shows the absorbed VHE $\gamma$-ray spectrum assuming an intrinsic $\Gamma=-1.5$ (displayed in black).

We find that the EBL density does not change much if the encountering galaxy has a stellar mass less than $10^{8} \rm ~ M_\odot$, even for a very close encounter (e.g. as close as $\sim 20 \rm ~ kpc$). Whereas if the galaxy has stellar mass greater than $10^{9} \rm ~ M_\odot$ the impact can be significant depending on the distance from the line-of-sight. Large galaxies with stellar masses $\sim 10^{11} \rm ~ M_\odot$  have substantial effect on the spectrum even at relatively large distances (e.g. $\sim 80\rm ~ kpc$) from the line-of-sight.  However, the impact on the opacity can only be observed if the galaxy's stellar mass is higher than $10^{10} \rm ~ M_\odot$.

Table \ref{tab:gam} shows the spectral index ($\Gamma$) fitted at  two $\gamma$-ray ($E_\gamma$) energy ranges ($0.01<E_\gamma \leq 0.33\rm ~ TeV$ and $0.33 < E_\gamma \leq 10.8\rm ~ TeV$) for different galaxy masses and distances. The last row in the table shows the spectral index for this model where there is no galaxy inserted near to the $\gamma$-ray path. The effects of a galaxy encounter with the $\gamma$-ray path is significant only on the highest energy range compared to the lower energy range.

A steepening of the spectral index $\Gamma$ due to a galaxy located close to the $\gamma$-ray path can only be seen if the galaxy's stellar mass is larger than $10^{10} \rm ~ M_\odot$. The significance of the effect varies with the $\gamma$-ray energy: for $E_\gamma < 0.33 \rm ~  TeV$, there is no observable change in $\Gamma$; for the $0.33 < E_\gamma \leq 10.8 \rm ~ TeV$ range a galaxy with stellar mass $5 \times 10^{11} \rm ~ M_\odot$ causes a clearly visible steepening of $\gamma$-ray spectrum if it is close to the line-of-sight ($\lesssim 40 \rm ~ kpc$).  Although the probability of finding galaxies with stellar mass of $> 10^{12} \rm ~  M_\odot$ is extremely low, they could have an enormous effect on the $\gamma$-ray spectrum even if they are located as far as $\sim 80 \rm ~ kpc$ form the line-of-sight.  The scatter in the observed spectral indices $\Gamma$ of the VHE $\gamma$-ray sources displayed in Fig. \ref{fig:slo} cannot be caused by galaxy proximity to the $\gamma$-ray paths as they are too infrequent to cause that amount of scatter observed.

\begin{table}
\caption{The spectral index $\Gamma$s for different masses of galaxies at different distances}
\label{tab:gam}
\scalebox{0.8}{
\begin{tabular}{||l||l|l|l|l|l||}
\hline
Mass [$M_\odot$]                 & Energy [TeV]                  & $\Gamma_{20\,\text{kpc}}$ & $\Gamma_{40\,\text{kpc}}$ & $\Gamma_{60\,\text{kpc}}$ & $\Gamma_{80\,\text{kpc}}$ \\
\hline
\hline                     
\multirow{2}{*}{$5\times10^{6}$} & $0.01\leq E_\gamma \leq 0.33$ & 1.50   & 1.50   & 1.50   & 1.50     \\\cline{2-6}
                                 & $0.33<E_\gamma \leq 10.8$     & 1.53   & 1.53   & 1.53   & 1.53     \\
\hline
\hline
\multirow{2}{*}{$5\times10^{7}$} & $0.01\leq E_\gamma\leq 0.33$  & 1.50   & 1.50   & 1.50   & 1.50     \\\cline{2-6}
                                 & $0.33<E_\gamma \leq 10.8$     & 1.53   & 1.53   & 1.53   & 1.53     \\
\hline
\hline
\multirow{2}{*}{$5\times10^{8}$} & $0.01\leq E_\gamma\leq 0.33$  & 1.50   & 1.50   & 1.50   & 1.50     \\\cline{2-6}
                                 & $0.33<E_\gamma \leq 10.8$     & 1.53   & 1.53   & 1.53   & 1.53     \\
\hline
\hline
\multirow{2}{*}{$5\times10^{9}$} & $0.01\leq E_\gamma\leq 0.33$  & 1.50  & 1.50    & 1.50   & 1.50     \\\cline{2-6}
                                 & $0.33<E_\gamma \leq 10.8$     & 1.53  & 1.53    & 1.53   & 1.53     \\
\hline
\hline
\multirow{2}{*}{$5\times10^{10}$}& $0.01\leq E_\gamma\leq 0.33$  & 1.50  & 1.50    & 1.50   & 1.50     \\\cline{2-6}
                                 & $0.33<E_\gamma \leq 10.8$     & 1.54  & 1.53    & 1.53   & 1.53     \\
\hline
\hline
\multirow{2}{*}{$5\times10^{11}$}& $0.01\leq E_\gamma\leq 0.33$  & 1.50  & 1.50    & 1.50   & 1.50     \\\cline{2-6}
                                 & $0.33<E_\gamma \leq 10.8$     & 1.62  & 1.55    & 1.54   & 1.54     \\
\hline
\multirow{2}{*}{$5\times10^{12}$}& $0.01\leq E_\gamma\leq 0.33$  & 1.53  & 1.51    & 1.50   & 1.50     \\\cline{2-6}
                                 & $0.33<E_\gamma \leq 10.8$     & 2.39  & 1.75    & 1.63   & 1.59     \\
\hline
\hline           
\multirow{2}{*}{No galaxy}       & $0.01\leq E_\gamma\leq 0.33$ &\multicolumn{4}{c}{1.50}              \\\cline{2-6}
                                 & $0.33<E_\gamma \leq 10.8$    &\multicolumn{4}{c}{1.53}              \\
\hline
\hline                     
\end{tabular}
}
\end{table}
\section{Summary and conclusion}
\label{sec:con}
With this work we introduce a new hybrid EBL model combining semi-analytic and forward evolution approaches to determine the EBL and its fluctuations in a realistic manner. Instances of the local EBL are calculated at the centre of a $50 ~ h^{-1} \rm Mpc$ spheres as the cumulative intensity of all galaxies within those spheres. The galaxy properties are obtained from the MR7 galaxy catalogue. The spectra of individual galaxies are generated using those galaxy properties and the stellar population synthesis spectral library BC03. The EBL originating from all sources outside the $50 ~ h^{-1} \rm Mpc$ spheres is assumed to be homogeneous and calculated analytically following KF17. We find one order of magnitude fluctuations in the local EBL. However, when we compute the total EBL by adding the homogeneous background contribution the variation is less than $\sim 1 \%$. This is expected as the local EBL contribution is only a small fraction of the total EBL intensity.

Based on the hybrid model we calculate the EBL along $\sim 1000$ $\gamma$-ray paths up to $z=0.5$, i.e., enveloping the $\gamma$-ray we construct light cone cylinders of $50\,h^{-1}\rm ~ Mpc$ comoving radius through the MR7 simulation volume within which we compute the local EBL based on the local galaxy content. The EBL beyond the cylinder is determined analytically for every time step. The hybrid EBL model approach then allows us to integrate the VHE $\gamma$-ray absorption along the $\gamma$-ray paths and determine the change of the observed $\gamma$-ray spectral index due to the preferential absorption of VHE $\gamma$-rays by the EBL photons.

The integrated opacities agree with the observed steepening of the hard $\gamma$-ray spectra as a function of redshift. But, we find only negligible variations of the opacities along the different $\gamma$-ray paths. Thus, our hybrid model which takes into account the contribution of individual galaxies close to the $\gamma$-ray path cannot account for the scatter in the spectral index seen in Fig. \ref{fig:slo}.

To cover very close encounters of galaxies with $\gamma$-ray paths (which are not apparent in our statistical sample) we manually insert an individual galaxy of different stellar mass ($5\times10^6 \rm ~ M_\odot$ to $5\times10^{12} \rm ~ M_\odot$) very close to the line-of-sight. We then compute $\gamma$-ray absorption and the increase of the spectral index $\Gamma$ which is set to be $\Gamma=-1.5$ without absorption. We find that the presence of a single galaxy with a stellar mass $<10^9 \rm ~ M_\odot$ has no measurable impact on $\Gamma$ independent of its distance to the line-of-sight. On the other hand, high mass galaxies with $10^{10}-10^{12} \rm ~ M_\odot$ at distances less than $50 \rm ~  kpc$ from the line-of-sight cause a clearly visible steepening of the VHE $\gamma$-ray spectrum. Given the future increase of the number of observed VHE $\gamma$-ray sources (roughly a factor of $10$  with the coming CTA) those encounters may become more important.


\section*{Acknowledgments}
We thank the anonymous referee for providing constructive comments. This research was supported by the German academic exchange service (DAAD). The Millennium Simulation (MR7) databases used in this paper and the web application providing online access to them were constructed as part of the activities of the German Astrophysical Virtual Observatory (GAVO). Finally, we thank Dr. N. Shafi and Dr. P. Marchegiani for useful comments.

\bibliographystyle{mn2e}
\bibliography{Bibliography}

\label{lastpage}
\end{document}